\documentclass[conference]{IEEEtran}
\IEEEoverridecommandlockouts
\usepackage{cite}
\usepackage{amsmath,amssymb,amsfonts}
\usepackage{algorithmic}
\usepackage{booktabs}
\usepackage{graphicx}
\usepackage{textcomp}
\usepackage{svg}
\usepackage{xcolor}
\usepackage{colortbl}
\usepackage{eso-pic}
\usepackage{multirow}
\usepackage{array}
\usepackage{threeparttable}
\usepackage{url}
\usepackage{float}
\usepackage{balance}

\newcolumntype{C}{>{\centering}p{0.045\textwidth}}
\newcolumntype{X}{>{\centering\arraybackslash}m{0.045\textwidth}}

\def\BibTeX{{\rm B\kern-.05em{\sc i\kern-.025em b}\kern-.08em
    T\kern-.1667em\lower.7ex\hbox{E}\kern-.125emX}}
    
\usepackage[hyperfootnotes=false]{hyperref}

\newcommand\copyrighttext{%
  \footnotesize \textcopyright 2023 IEEE. Personal use of this material is permitted.
  Permission from IEEE must be obtained for all other uses, in any current or future
  media, including reprinting/republishing this material for advertising or promotional
  purposes, creating new collective works, for resale or redistribution to servers or
  lists, or reuse of any copyrighted component of this work in other works.
  DOI: \href{https://doi.org/10.1109/IC2E59103.2023.00030}{https://doi.org/10.1109/IC2E59103.2023.00030}}
\newcommand\copyrightnotice{%
\AddToShipoutPicture*{%
\put(45,30){%
\centering
\fbox{\parbox{\dimexpr\textwidth-\fboxsep-\fboxrule\relax}{\copyrighttext}}
}}}

\begin{document}

\title{Evaluation of Data Enrichment Methods for Distributed Stream Processing Systems}

\author{
\IEEEauthorblockN{Dominik Scheinert,
Fabian Casares,
Morgan K. Geldenhuys,
Kevin Styp-Rekowski,
and Odej Kao}
\IEEEauthorblockA{Technische Universit{\"a}t Berlin, Germany, \{firstname.lastname\}@tu-berlin.de}
}

\maketitle
\copyrightnotice

\begin{abstract}
Stream processing has become a critical component in the architecture of modern applications. 
With the exponential growth of data generation from sources such as the Internet of Things, business intelligence, and telecommunications, real-time processing of unbounded data streams has become a necessity. 
DSP systems provide a solution to this challenge, offering high horizontal scalability, fault-tolerant execution, and the ability to process data streams from multiple sources in a single DSP job.
Often enough though, data streams need to be enriched with extra information for correct processing, which introduces additional dependencies and potential bottlenecks.

In this paper, we present an in-depth evaluation of data enrichment methods for DSP systems and identify the different use cases for stream processing in modern systems. 
Using a representative DSP system and conducting the evaluation in a realistic cloud environment, we found that outsourcing enrichment data to the DSP system can improve performance for specific use cases. 
However, this increased resource consumption highlights the need for stream processing solutions specifically designed for the performance-intensive workloads of cloud-based applications.

\end{abstract}


\section{Introduction}
\label{sec:introduction}

Continuous generation of data from web applications and connected devices characterizes modern-day technology and commonly leads to large volumes of data and the need for their processing. 
Usually, data processing results are of immediate interest to subsequent actions, which is why near-to real-time processing is often desirable. 
Many areas such as Internet of Things, operational business intelligence, and fraud detection rely on near-to real-time processing~\cite{IsahAMAZK19,NasiriNG19} of data generated from highly distributed systems. 
However, to obtain a meaningful analysis of data from disparate sources, the respective data traditionally need to be copied into a high-capacity data store for analysis, which can be complex and negatively impact performance. 
Classical batch processing is no longer sufficient for many use cases, hence, the development and thoughtful usage of Distributed Stream Processing (DSP) systems~\cite{KulkarniBFKKMPR15,ZahariaXWDADMRV16,CarboneKEMHT15,ToshniwalTSRPKJGFDBMR14,GulenkoA0BK20}, optimized to process continuous streams of data on a large scale, is important.

DSP systems are optimized to process large streams of continuous data.
Importantly, their ability to offer high throughput processing rates makes them suitable for many use cases.
With these systems, near to real-time processing is mandatory, and the distributed execution of DSP jobs in the cloud allows for the cluster to be dynamically scaled up and down~\cite{KalavriLHDFR18,GeldenhuysSKT22,PfisterLNPGSGT21,GedikSHW14} and hence gracefully adapt to changing workloads~\cite{GontarskaGSWPT21,HuKZ19,KalimCWLWLFQLCW19} or optimized toward desired objectives~\cite{GeldenhuysPSTK22,GeldenhuysTK20,JayasekaraHK20,GeldenhuysTGLK19,FloratouAGRR17}. 
Recent DSP frameworks also offer stateful stream processing, which allows operators to store and access intermediate data within the cluster, making more complex processing of stream events possible, and guaranteeing exactly-once semantics. 
In many cases though, dependencies exist between heterogeneous workloads and the processes which consume them, which may result in latency issues as well as scaling bottlenecks, for instance, if external systems must be accessed during execution, or a performance-heavy workload must be executed within the DSP job of interest.

Related works have addressed the enrichment of events in a DSP system through external databases~\cite{DerakhshanSS13,JeonLK19,KimL20c}, evaluated the performance of running a Machine Learning (ML) model embedded in a DSP system~\cite{HorchidanKKC22}, and also the unification of stream and batch jobs in a single application~\cite{GarefalakisKP19,MeldrumSKC0H19}. 
However, specific use case categories for stream data enrichment are yet to be identified, and specific enrichment methods for these categories need to be presented and evaluated. 

This paper aims to address the issue of streaming data enrichment for DSP systems by providing an evaluation of different enrichment methods and identifying the proper enrichment method for a given use case. 
We primarily focus on latency-critical applications, and select Apache Flink as a representative DSP system for our evaluation due to its wide usage by big corporations on a large scale~\cite{CarboneEFHRT17}.

\textit{Contributions.} The contributions of this paper are:

\begin{itemize}
    \item Problem analysis and investigation of assumptions and common use cases, ultimately leading to the definition of general use case categories.
    \item Detailed empirical evaluation of various data enrichment methods in combination with different representative use cases, providing a better understanding of situation-dependent applicability.
    \item Openly available repository\footnote{\url{https://github.com/dos-group/stream-processing-enrichment-methods}} with all relevant experiment-related artifacts. We provide comprehensive documentation and examples for reproducing our setup.
\end{itemize}

\autoref{sec:problem_analysis} conducts a problem analysis, thereby identifying the assumptions and use cases for performing data enrichment for DSP systems. \autoref{sec:experiment_methodology} presents the data enrichment methods, infrastructure setup, selected use cases, and evaluation metrics. \autoref{sec:experiment_results} presents and discusses our results. \autoref{sec:related_work} describes the related work on data enrichment strategies, while \autoref{sec:conclusion} concludes the paper.

\section{Problem Analysis}
\label{sec:problem_analysis}
In this section, we first present our assumptions regarding data enrichment in DSP systems and then elaborate on general applicable use cases in this field.

\subsection{Assumptions}
The processing of unbounded data streams requires DSP systems to in theory execute indefinitely. 
As the near-to real-time processing of events is crucial for a wide range of applications, various requirements must be met by the DSP system depending on the application.
In this work, we therefore primarily focus on low-latency streaming jobs, but there are other relevant aspects that must be considered.

One such aspect is the reliability of a streaming job. 
System failures are common in large clusters, and for DSP jobs that are required to operate indefinitely, failures are inevitable, making it essential for the DSP system to guarantee exactly-once semantics and recovery from failures. 
Data enrichment methods should consequently take this into account and exhibit certain robustness.
Another important requirement is scalability, as the workload of a stream may change over time, leading to the need for adding new resources to avoid performance degradation or removing resources to optimize resource utilization.
Depending on the concrete design, this can also affect an employed data enrichment method.
Lastly, streaming applications and their underlying architectures can quickly lead to an increase in complexity, i.e., in light of the distributed execution graph, heterogeneous data sources, or data sinks.
This is further reinforced through support for libraries that enable ML or graph processing.

\subsection{Data Enrichment Use Cases}

Data processing in a DSP system often requires additional context for accurate analysis and interpretation. 
To achieve this, enrichment with additional data can be performed during the execution of the DSP job.
There are several reasons why data enrichment could be necessary. 
Firstly, the incoming data streams may lack the necessary information to provide accurate insights. 
For example, if a system is monitoring sensor data, it may be necessary to enrich the data with information about the location, time, or weather conditions to understand the context in which the data was collected.
In case of constrained network links as in IoT environments, this can furthermore reduce the size of the individual events and lower overall network overhead, as events remain compact in size until they reach the DSP system in the cloud, where they are eventually enriched.
Secondly, data enrichment can help to detect anomalies or patterns that may be hidden in the data. 
By adding more information to the data, it may be possible to identify patterns that were not apparent before, such as identifying fraud or predicting a failure before it occurs.
Thirdly, data enrichment can help to integrate data from multiple sources. 
In DSP systems, data may come from multiple sources, and integrating this data can be a complex process. 
By enriching the data with additional information, it may be possible to integrate data from different sources and provide a more complete picture of the target system.

Data enrichment during execution can vary greatly based on the use case, and the underlying architecture and priorities can be unique. 
Due to the diversity of use cases, there is no one-size-fits-all solution for enriching events in a DSP system. 
In order to carve out the advantages and disadvantages of particular solutions, we conduct a comprehensive evaluation of different enrichment methods.
For this endeavor, we derive the following broad use case categories, along the criteria of data availability, data volume, and time sensitivity:

\begin{itemize}
\item \underline{Simple Queries:} In many instances, a data enrichment use case includes common operations, for instance, a simple key-value database query, where the event contains a key and there is one value in the database that can be efficiently queried.
An example would be the location of a sensor in an IoT environment that could be queried by a key.
Other examples include API requests or inference services of ML applications.
We envision that for this category, all these examples have in common that the response time is fairly constant, yet loading the entirety of information is not possible, for example, due to capacity limitations (e.g. databases), service barriers (APIs of closed systems), or undeterministic data (e.g. time-sensitive information such as weather data).
\item \underline{Complex Queries:} This category resembles the first one, with the difference that queries or data formats are more complex, and hence response times can be fluctuating.
This is for instance the case for complex database queries that include multiple join statements (e.g. for fraud detection), or for API requests which trigger different behavior depending on the payload.
This increased complexity hinders the migration of information directly to the respective DSP system for accelerated data enrichment.
\item \underline{Finite Data Sources:} In certain scenarios, the information used for enriching streaming events might be compact in size and hence might qualify for a migration directly to the DSP system as embedded state. 
Examples range from small disclosed ML models, which generate a data output for each data input, to static sensor information.
While potentially beneficial for event latencies, additional challenges are raised with respect to state handling within the DSP system as well as resource management. 
\end{itemize}

The goal of our evaluation is to determine suitable enrichment methods for specific use cases originating from our defined use case categories, to allow for guidance, and to enable practitioners to make informed decisions.

\subsection{Enrichment Methods}

In the following, we discuss various methods of enriching events in state-of-the-art DSP systems. 
These methods serve as a baseline for investigating the previously identified categories of data enrichment use cases.\\
\textbf{Datasource Client.} This method connects to an external data source to access the data for single / batches of events, which can be performed either synchronously or asynchronously.
\begin{itemize}
    \item \underline{Synchronous}: A synchronous client is the simplest way to connect to an external data source. This method enriches each event with the result of a synchronous query to the data source. Although a blocking procedure, the advantage of this method is that it can be easily integrated into existing architectures, and most conventional databases provide a synchronous client. If a pattern recognition model is used for enrichment, this method can be applied if the model is executed in an external service.
    \item \underline{Asynchronous}: This method involves using an asynchronous client to connect to the external data source, allowing for parallel execution of queries and improved utilization of query times. This requires the availability of an asynchronous client library. If no such library is available, asynchronous queries can be simulated with a custom multi-threading implementation.
\end{itemize}
\textbf{Cache.} To reduce access to external and potentially slow data sources, a subset of the data can be cached for faster access and to reduce dependencies. The data format must be able to be cached, and the external data should not change frequently. For aggregation operations, caching can quickly become costly.
\begin{itemize}
    \item \underline{Local Caching}: This method caches a subset of the external data within the respective operation of the DSP system. In case of a cache miss, a query to the external data source is executed. This method reduces latency and the load on the external system, and can store non-serializable objects. The storage capacity of the local cache depends on the worker node's storage capacity.
    \item \underline{External In-Memory Database Cache}: This method caches a subset of the external data in an external in-memory database such as Redis. This creates an additional synchronous or asynchronous connection to the in-memory database, in addition to the connection to the disk-based data source. In case of a cache miss, an additional query to the disk-based data source must be executed. Although an entirely new system is additionally required, the advantage of this method is that the resources can be managed independently of the DSP system, allowing for caching of a larger amount of data. The external cache is also more transparent and modifiable, making it easier to keep it consistent with the disk-based database if necessary.
\end{itemize}
\textbf{Embedded State.} Caching methods maintain a connection to the external data source, leading to a direct dependency. To overcome this, this method involves loading the entire external data into the DSP system as a stream and treating it as another source. The events are then enriched by a join operation. This method requires that outsourcing external data is possible and that the amount of data is within the available resources of the DSP system. Modern systems such as Spark or Flink have an included state backend that can store large amounts of data using RocksDB. However, using a disk-based state in the DSP system can reduce performance. In-memory state is recommended for real-time processing, but requires a large amount of memory.
The embedded state can hence often be regarded as a special case of local caching.

These enrichment methods can be implemented in common DSP systems and serve as a foundation for our evaluation.
\section{Methodology}
\label{sec:experiment_methodology}

In this section, we present our experiment methodology by introducing the infrastructure we base our experiments on, the implemented use cases, and the utilized evaluation metrics.

\subsection{Infrastructure Setup}

\begin{figure}
  \centering
  \includegraphics[width=\columnwidth]{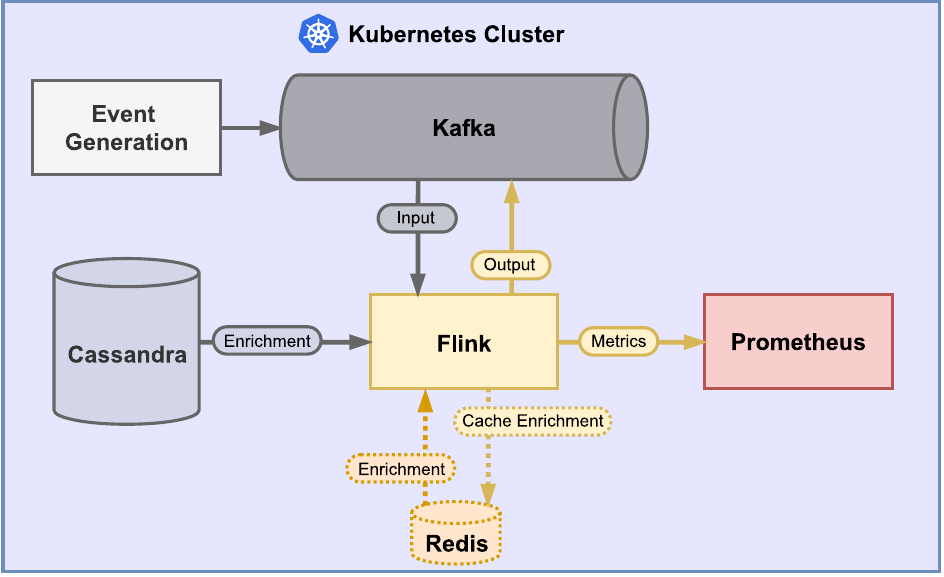}
  \caption{Visualization of infrastructure setup as well as interactions of all relevant components around the DSP system.}
  \label{fig:infrastructure}
\end{figure}

We select Apache Flink as a representative DSP system for our evaluation because it fulfills all previously defined assumptions like exactly-once semantics, fault
tolerance, horizontal scalability, near to real-time event processing with processing one
event at a time, and batch processing.
Our complete infrastructure around Flink, illustrated in~\autoref{fig:infrastructure}, is deployed in a Kubernetes cluster to ensure high availability, fast scaling, and ease of management. 
Apache Kafka is used as a messaging platform due to its high scalability, availability, and fault-tolerance. 
The different partitions of Kafka are evenly distributed among the sub-tasks of Flink. 
External data source enrichment is performed using Apache Cassandra, a column-based NoSQL database with high availability and fast read access. 
Redis is deployed as a cache and is used as a popular in-memory key-value NoSQL database. 
Prometheus, an open-source monitoring service with an integrated time series database, is deployed for a successful and meaningful evaluation. 
Flink metrics can be easily scraped periodically and stored in Prometheus for later access.

\begin{table}
\centering
\caption{Kubernetes Cluster Setup}
\begin{tabular}[t]{lp{0.58\linewidth}}
 \toprule
 Resource & Details\\
 \midrule
 Node - Machine & c2-standard-8 (32GB memory, 8 vCPUs\\
 & with 3.1GHz base frequency) \\ 
 Node - Disk & 100GB, pd-standard (backed by HDD) \\ 
 Software & Docker 20.10, Kubernetes 1.23, Java 11,\\
 & Flink 1.14, Kafka 2.8, Redis 7.0,\\
 & Cassandra 4.0, Prometheus 2.3, Python 3.7 \\
 \bottomrule
\end{tabular}
\label{table:experimental_setup}
\end{table}

In general, the deployment of the infrastructure was parameterized with the aim of conducting the largest possible number of experiments within the available financial resources, with some chosen parameters proving appropriate even via an exploratory approach.
The entire infrastructure was deployed in an 8-node Kubernetes cluster in a single data center of the Google Cloud Platform (GCP) using the Google Kubernetes Engine (GKE). 
Flink was deployed with a parallelism of 8, which is also true for the keyed window operators used in our streaming jobs to be presented, and each \textit{TaskManager} having 2 task slots. 
In contrast, Kafka was deployed with a cluster size of 3 and each topic with 8 partitions and a replication factor of 3. 
The Cassandra instance was deployed with replication factor 1 and the SimpleStrategy replication strategy. 
For monitoring the Flink metrics, a Prometheus instance was deployed on a single node.
In order to accurately simulate a real use case, Flink was always deployed on different nodes than the Cassandra instances using node selectors, since in a real use case the database is often kept separate, and deploying them on the same nodes can result in latency advantages.
Moreover, the event generation of the use cases was performed in a separate deployment in the Kubernetes cluster to simulate network overhead more realistically. 
\autoref{table:experimental_setup} summarizes the Kubernetes cluster setup and the used software.

\subsection{Use Case Implementations}

We implement two representative use cases, one for fraud detection and another one for log analytics.

\begin{figure}
 \begin{minipage}{\linewidth}
  \centering
  \includegraphics[width=\columnwidth]{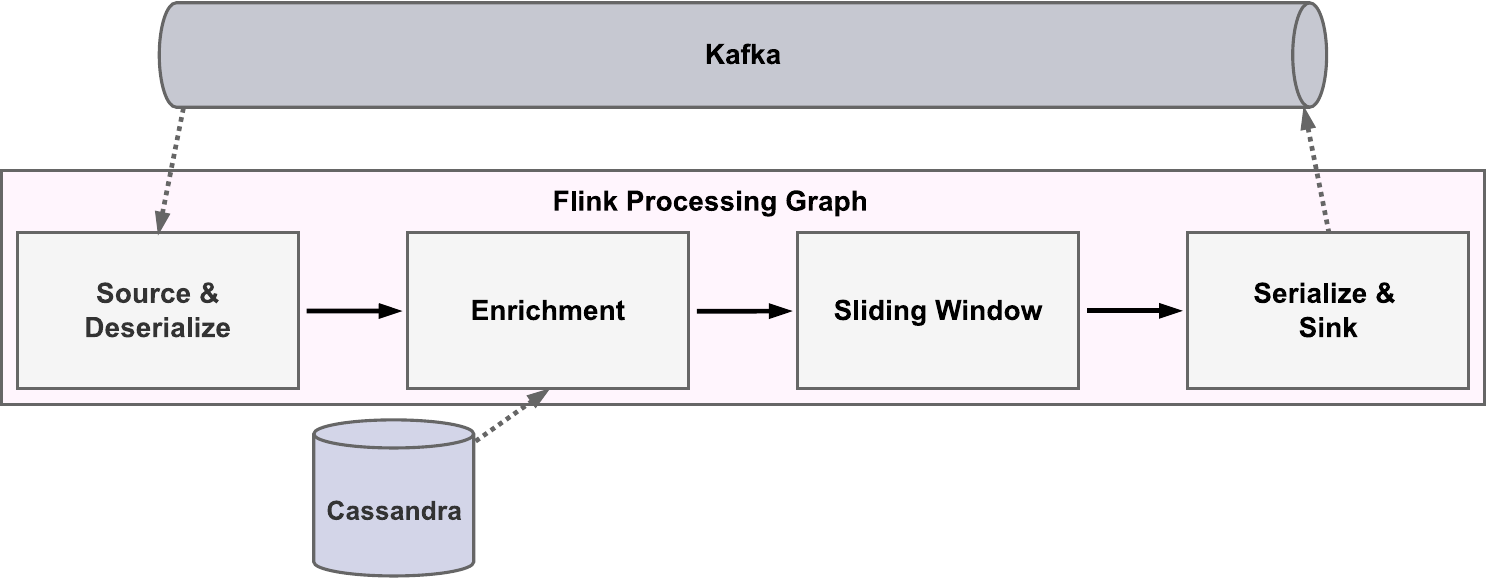}
  \caption{Fraud Detection: Various enrichment methods are put to the test.}
  \label{fig:fraud_detection_streaming_job}
 \end{minipage}
 
 \vspace{0.5cm}
 \begin{minipage}{\linewidth}
  \centering
  \includegraphics[width=.75\columnwidth]{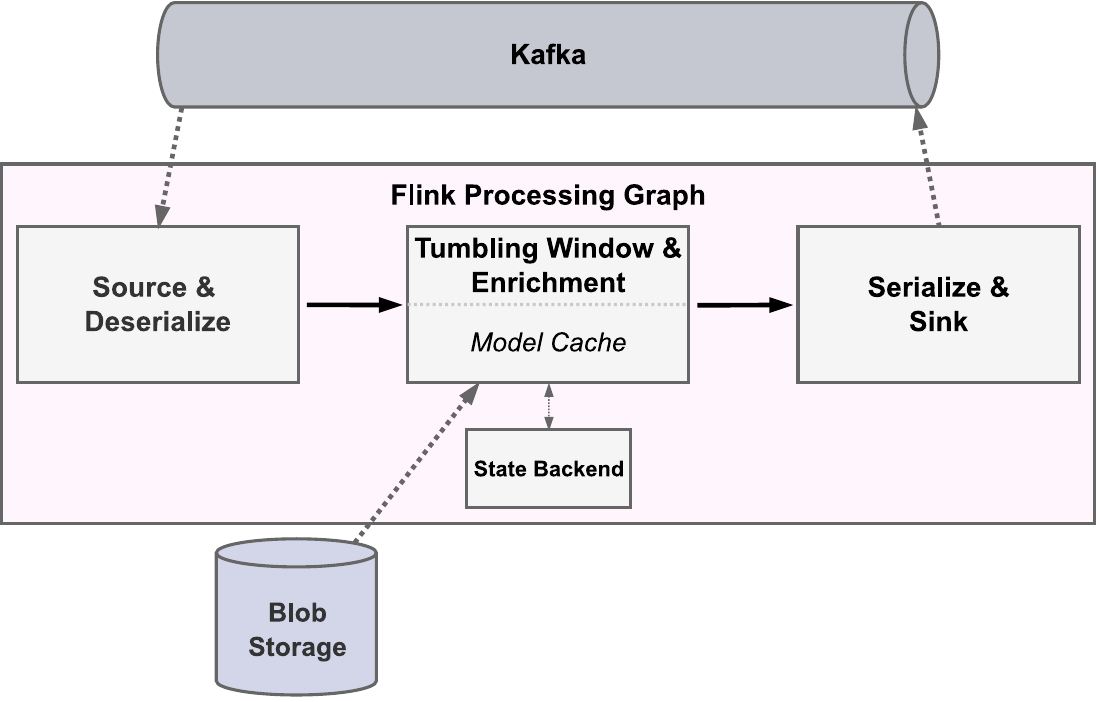}
  \caption{Log Analytics: ML models are loaded into the DSP system.}
  \label{fig:log_analysis_streaming_job}
 \end{minipage}
\end{figure}

For several reasons, \textbf{\underline{fraud detection}} is particularly well suited as a use case to evaluate different enrichment methods.
Historical data for enrichment in fraud detection is usually enormous and may require complex queries to get efficient results. 
The data structure can also be complex and outsourcing a subset or all of the data would be difficult. 
Additionally, the constant addition and editing of historical data may impact the enrichment process. 
The complexity of fraud detection highlights why different enrichment methods can be considered.
\textbf{Event Structure.} For this use case, we are generating credit card transaction events and writing them to Kafka. 
Each event contains three key pieces of information: transaction details, device information, and location information. 
The device and location information each have a hash field, which allows for unique identification. 
The transaction part can be identified by the account and receiver ID, and the transaction ID is unique across events. 
Parameters can be used to control the probability of events containing known 
information.\\
\textbf{Historical Data.} We store historical data in the form of transactions in a Cassandra database to enrich transaction events. 
Cassandra is a suitable choice because it can efficiently handle globally made transactions in a distributed setting, and its column-oriented schema is ideal for storing transaction events and querying them efficiently. Three separate tables were created to store device, location, and transaction information. 
The partition key for the device and location tables is the account ID, and the hash value is the cluster key. 
For the transaction table, the account ID is the partition key, and the receiver ID and transaction ID are both cluster keys.\\
\textbf{Streaming Job.} The job, illustrated in~\autoref{fig:fraud_detection_streaming_job}, reads transaction events from Kafka and enriches them with historical data. 
Depending on the enrichment method, either the external Cassandra instance is accessed or the events are enriched with outsourced data directly in Flink.
Since join operations are not possible with Cassandra, three different queries would thus be necessary.
The subsequent enrichment verifies if the recipient, device, or location has been used by the account before, and flags the event as suspicious if it has not. 
A sliding window operation is performed on the enriched data stream to analyze the transaction volume of an account in a certain period of time.
The resulting event includes the total transaction amount, the number of transactions in the window, and the suspicious flags. 
The event is then serialized and written back to Kafka.\\
\textbf{Enrichment Methods.} We applied various enrichment methods to the fraud detection use case in Flink. 
For the datasource client, two types of enrichment methods have been implemented: synchronous and asynchronous. 
The asynchronous method was implemented using the Async I/O API and the DataStax asynchronous client. 
For caching, three different methods were used: local caching with a \texttt{LinkedHashMap}, local caching with custom partition using the Flink \texttt{partitionCustom} function, and external caching using Redis, where the cache was executed in combination with the asynchronous Cassandra client in the \texttt{RichAsyncFunction} operation.
For embedded state, all historical data was stored in a Kafka topic as another source stream and joined with the latest transaction events. 

In contrast, our second use case refers exclusively to enrichment of streaming data with ML models, specifically log data.
\textbf{\underline{Log analytics}} is important for modern applications, as logs provide insight into user and system activity and can help identify errors and suspicious behavior. 
Our evaluation of enrichment of streaming events by ML models focuses on the effects of varying model sizes and numbers, assuming that different models are required for analysis. 
The models are executed embedded in Flink to avoid latency issues that could arise if the models were run externally in separate services. 
Running multiple embedded models is possible but requires careful management of memory usage to avoid crashes.\\
\textbf{Event Structure.} Log events are generated as input for the streaming job. 
An event contains a key that represents a specific service from which the log message originates,
another field for the content of the log message, and a timestamp.\\
\textbf{Model Source.} We assume that ML models will be stored in an external storage due to their high memory usage. 
This allows for efficient combination with other cloud-located services and easy access and updates of the models. 
However, for our use case, the external data source only plays a minor role, as the models are only fetched from the external data source once during initialization and then executed locally.\\
\textbf{Streaming Job.} A pre-trained ML model is used to make predictions within this job, as depicted in~\autoref{fig:log_analysis_streaming_job}. 
The log events are read from Kafka, then a Keyed Stream is created and a tumbling window operation is performed. 
The pre-trained model is loaded from Google Cloud Storage into Flink and processed with the ONNX Runtime, a high-performance engine for executing ML models that are compliant with the Open Neural Network Exchange (ONNX) format. 
The model is loaded into a ML library-specific session object to run the model with an input. 
The enrichment method loads the service-specific models once from the external data source and then stores them in the state backend for efficient access. 
A single result is generated per window, containing the service key, predictions, and window timestamps.

\subsection{Evaluation Metrics}

The evaluation primarily requires measurement of latency, which is not recommended to be obtained from Flink's built-in end-to-end latency metric as it impacts the cluster's performance and, in the worst case scenario, only reflects the queue time for events traversing window operators. 
Thus, a custom latency metric was created for the fraud detection use case, which considers the waiting time, only records the first occurrence of an event in a sliding window, and stores the final latency in a Flink metric histogram. 
The metric can then be accessed in Prometheus along with other built-in Flink metrics.
For the log analytics use case such custom solution is not necessary, which is why the latency metric is obtained using a regular histogram.
In addition to latencies, we also consider the consumption rate as well as system load where feasible.
\section{Experiment Results}
\label{sec:experiment_results}

This section presents our experiments and associated results, and discusses our key findings.
To obtain a meaningful evaluation, each experiment was performed three times. 
The plots of the results always include the data from all three executions, whereas the mean of the three data points is highlighted. 

\subsection{Fraud Detection Use Case}

The fraud detection use case uses the default configuration of the \texttt{HashMapStateBackend} with a checkpointing interval of 10 minutes to avoid affecting performance as would be the case when using a performance-intensive state backend such as the \texttt{EmbeddedRocksDBStateBackend}. A window size of 10 seconds and a slide size of 5 seconds were chosen for the sliding window operation, balancing the detection of fraud and performance requirements. The chosen parameters ensure that no large state is required during execution, which could affect latency and distort results.

\subsubsection{Datasource Clients}

\begin{figure}
 \begin{minipage}{\linewidth}
  \centering
  \includegraphics[width=\columnwidth]{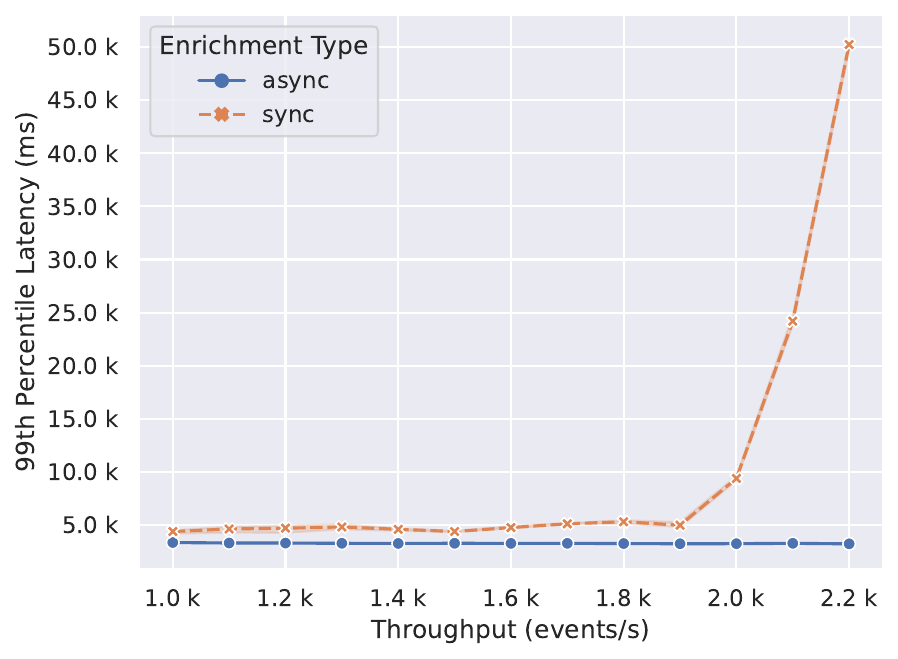}
    \caption{Latencies of data enrichment by \textbf{synchronous} and \textbf{asynchronous} Cassandra queries. Every 5 minutes, the throughput increases by 100 events/s and each data point corresponds to the average latency of all sub-tasks.}
  \label{fig:sync_async_latency}
 \end{minipage}
 
 \vspace{0.25cm}
 \begin{minipage}{\linewidth}
  \centering
    \includegraphics[width=\columnwidth]{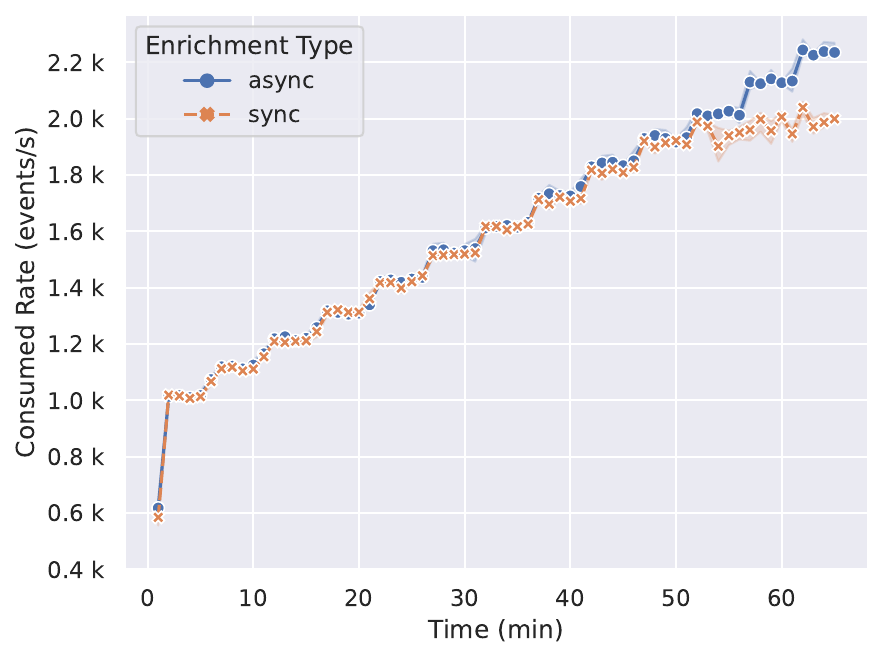}
    \caption{Comparing the actual consumption rate in events/s of the \textbf{synchronous} and \textbf{asynchronous} enrichment streaming jobs. The generated events/s are increased by 100 event/s every 5 minutes, which is also reflected in the consumption rate of both enrichment methods until minute 50.}
    \label{fig:sync_async_consumedrate}
 \end{minipage}
\end{figure}

\begin{figure}
 \begin{minipage}{\linewidth}
  \centering
    \includegraphics[width=\columnwidth]{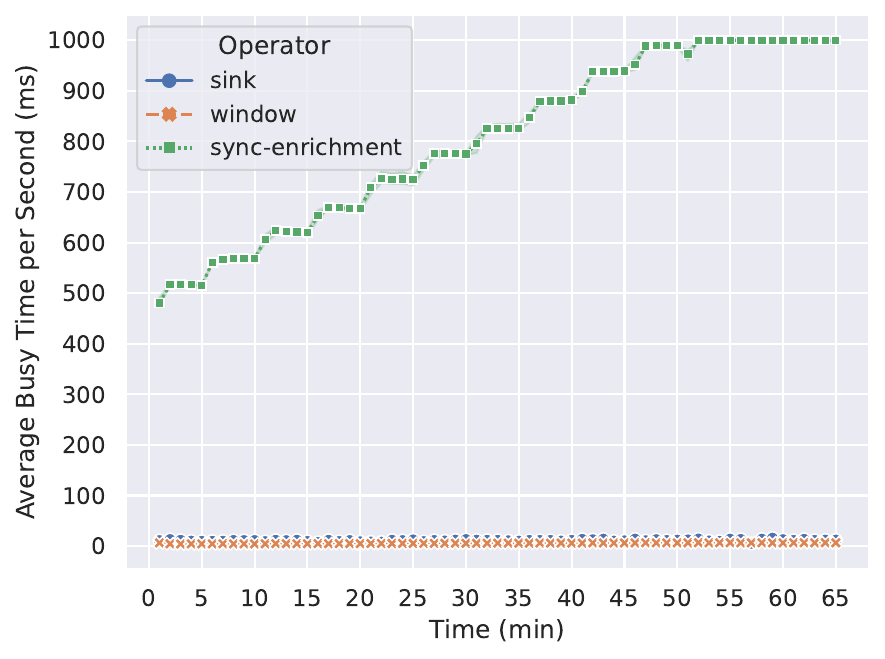}
    \caption{The busy values of the streaming job with \textbf{synchronous} enrichment as throughput increases. The enrichment tasks clearly take up the highest load even with the lowest generated throughput (100\% load for 1,900 events/s).}
    \label{fig:sync_busy_rates}
 \end{minipage}
 
 \vspace{0.25cm}
 \begin{minipage}{\linewidth}
    \centering
    \includegraphics[width=\columnwidth]{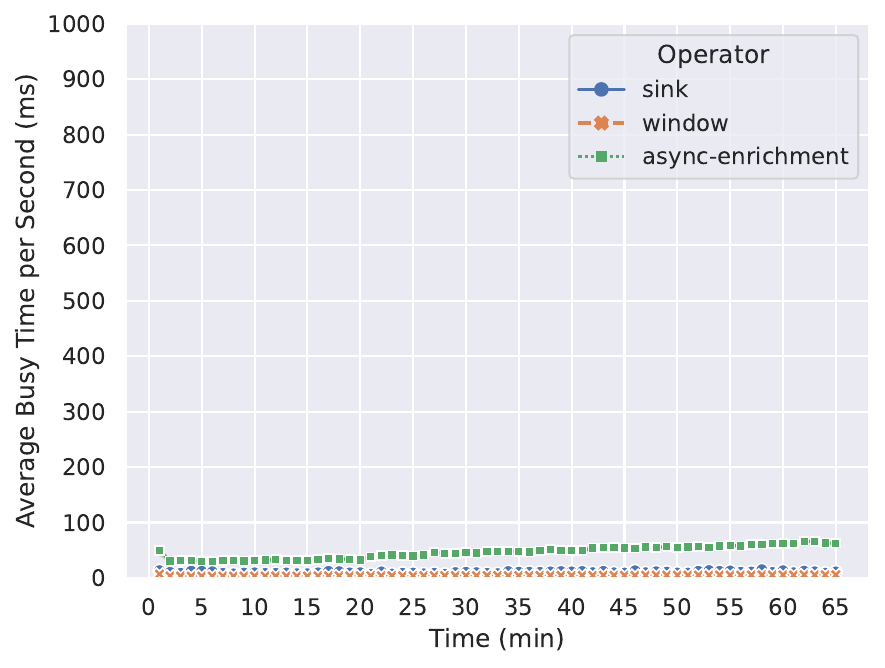}
    \caption{The busy values of the execution with \textbf{asynchronous} enrichment are shown with increasing generated throughput. It can be seen that the enrichment tasks have the highest busy values compared to the other tasks, but the tasks never exceed 100ms per second of busy time (less than 10\% load).}
    \label{fig:async_busy_rates}
 \end{minipage}
\end{figure}

The purpose of the first experiment was to compare the latency and throughput of synchronous and asynchronous Cassandra clients under different conditions. 
The experiment was conducted using throughput rates ranging from 1,000 to 2,200 events per second in increments of 100 events per second. 
The results, illustrated in~\autoref{fig:sync_async_latency}, showed that asynchronous enrichment had a fairly constant latency that was always lower than the latency of enrichment by synchronous queries. 
Synchronous enrichment showed a slight increase in latency from 1,600 events/s, and a significant increase from 1,900 events/s, reaching a latency of approximately 50 seconds at 2,200 events/s. 
This rapid increase in latency is due to the maximum throughput being reached, leading to the back pressure mechanism taking effect, delaying event processing. 
The comparison of the consumed rate of the two streaming jobs was measured using the Kafka metric \textit{records-consumed-rate} and is depicted in~\autoref{fig:sync_async_consumedrate}. 
The data indicated that synchronous enrichment consumes slightly fewer events per second than asynchronous enrichment. 
Synchronous enrichment had a maximum throughput of approximately 1,900 events/s and reached a load of 100\% after 50 minutes (~\autoref{fig:sync_busy_rates}). 
On the other hand, asynchronous enrichment had a lower load, and the busy values of the enrichment tasks only increased slightly and remained below 100ms per second (~\autoref{fig:async_busy_rates}). 
In the streaming job with synchronous enrichment, the enrichment task had the highest load and reached 100\% at 1,900 events/s.
In conclusion, the results showcase the limitations of synchronous enrichment, and while asynchronous enrichment comes out superior, it will as well face problems in light of substantially higher throughput rates due to its technical implications, underlining the need for caching strategies, as discussed next.

\subsubsection{Caching Methods}

\begin{figure}
 \begin{minipage}{\linewidth}
    \centering
    \includegraphics[width=\columnwidth]{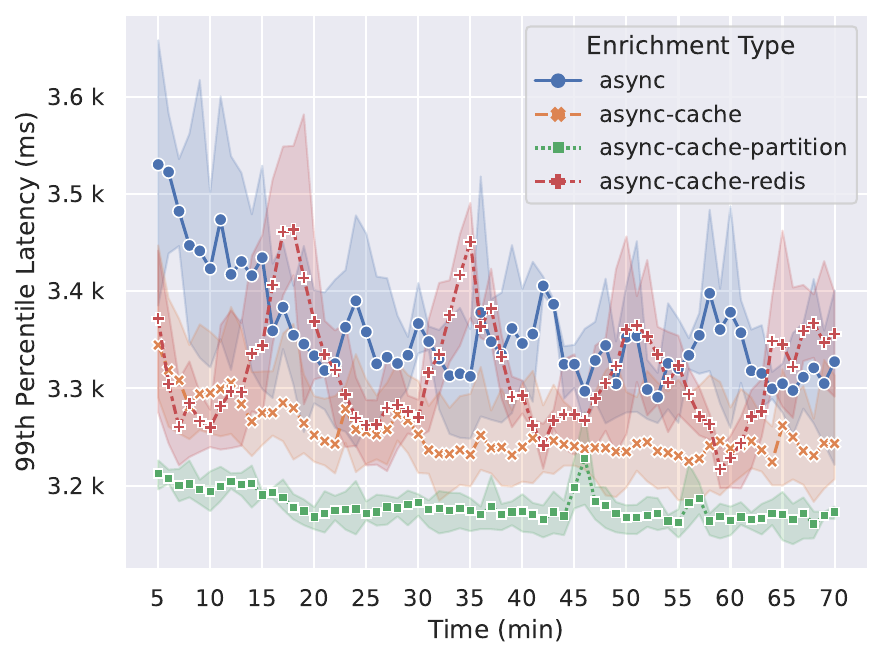}
    \caption{Latency comparison of different cache-based enrichment methods. Using a preceding custom partitioner evidently yields the most stable latencies.}
    \label{fig:caches_latency}
 \end{minipage}
 
 \vspace{0.25cm}
 \begin{minipage}{\linewidth}
    \centering
    \includegraphics[width=\columnwidth]{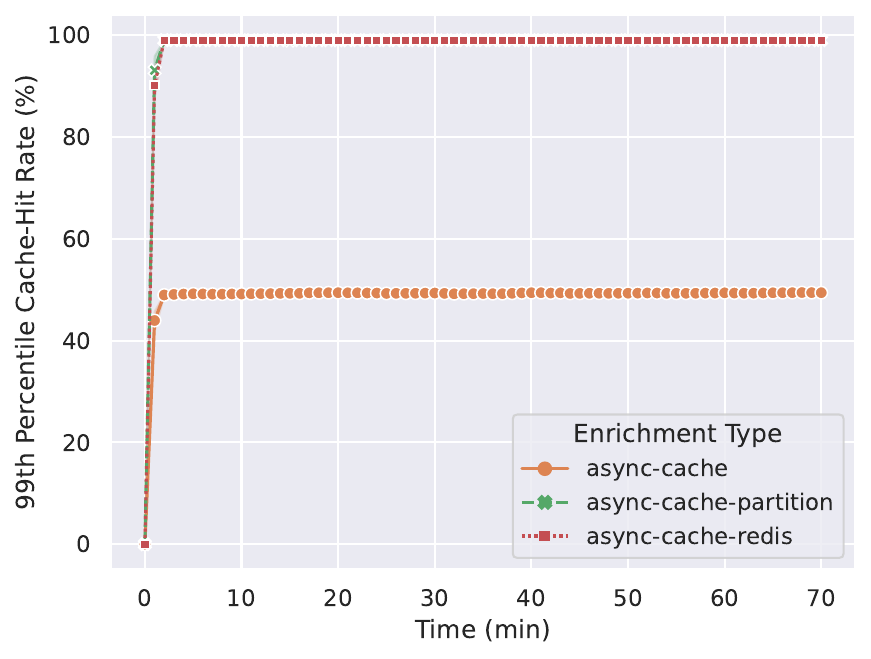}
    \caption{Cache hit rates corresponding to~\autoref{fig:caches_latency}. When caching with a preceding custom partitioner, each key is assigned to a specific task and thus each cache entry exists exactly once in all local caches, leading to a 100\% cache-hit rate in this evaluation as cache entries are filled without redundancy.}
    \label{fig:caches_cachehit_rates}
 \end{minipage}
\end{figure}

\begin{figure}
 \begin{minipage}{\linewidth}
    \centering
    \includegraphics[width=\columnwidth]{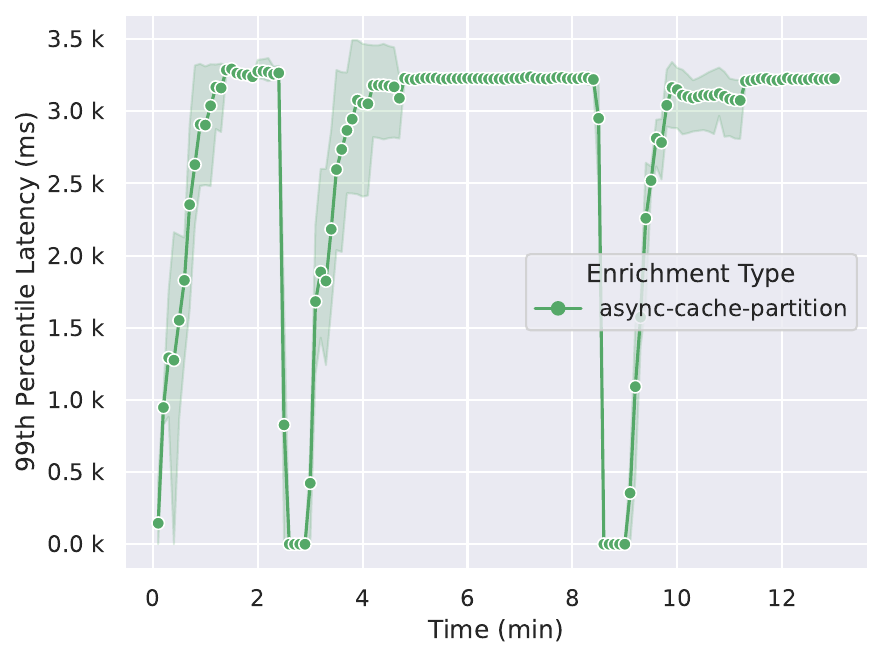}
    \caption{Cache-based enrichment with preceding custom partitioner in light of all \textit{TaskManagers} failing at once. Stable latencies are restored once the caches are refilled. Latencies smaller than the stable latencies do not reflect reality but are the result of Prometheus metric aggregations over time, which are temporarily corrupted after the complete failure of all \textit{TaskManagers}.}
    \label{fig:cache_partition_failures_latency}
 \end{minipage}
 
 \vspace{0.25cm}
 \begin{minipage}{\linewidth}
    \centering
    \includegraphics[width=\columnwidth]{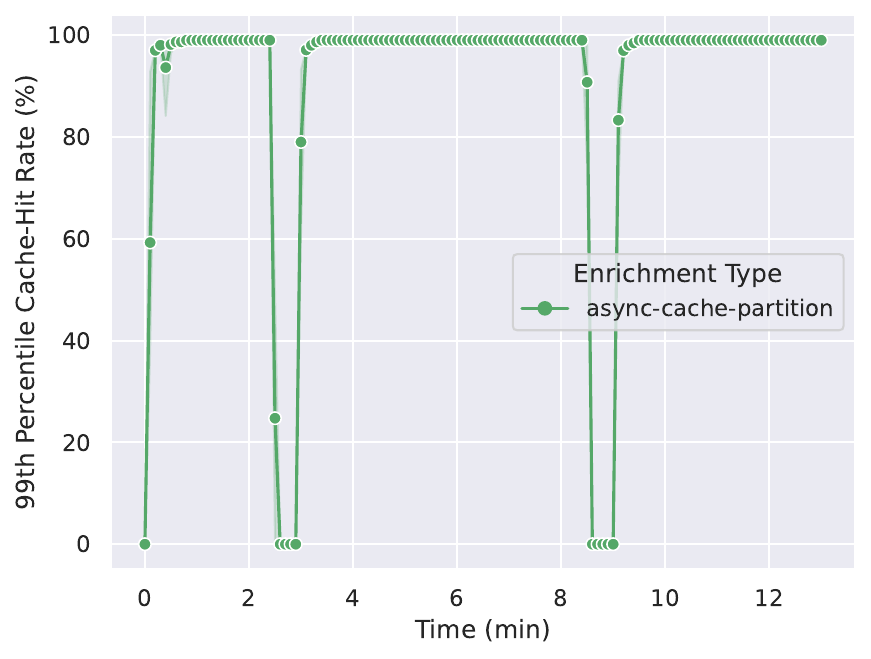}
    \caption{Cache-hit rates corresponding to~\autoref{fig:cache_partition_failures_latency}. After the restart of all \textit{TaskManagers}, the respective local caches are flushed, which is why cache-hit rates drop to 0\% and then quickly recover over time.}
    \label{fig:cache_partition_failures_cachehit}
 \end{minipage}
\end{figure}

In the previous experiments, it has become clear that asynchronous enrichment allows both better latency and higher throughput than enrichment using synchronous database queries.
We consequently now evaluate enrichment methods that use a cache to store database records combined with asynchronous Cassandra queries. 
The evaluation was conducted under two factors - the generation of events and the size of caches. 
To maintain the uniformity of the evaluation, each transaction event was required to be unique and generated uniformly. 
Also, the number of cache entries had to be equal to the amount of entries of the external cache. 
A total maximum number of cache entries was defined, with the sum of all cache entries being equal to 24,000. 
A fixed throughput of 4,000 events per second was selected for the latency evaluation, with each execution running for 70 minutes. 
The results, depicted in~\autoref{fig:caches_latency}, indicate that caching using a preceding custom partitioner has the best latency and cache hit rate of 100\% (~\autoref{fig:caches_cachehit_rates}). 
With our particular experiment design, local caching without a preceding custom partitioner achieved a cache hit rate of only 50\% and lower latency than asynchronous enrichment without cache but worse latency than caching with a preceding custom partitioner.
The local cache size for each local cache was 3,000. 
In contrast, asynchronous enrichment without cache showed a more volatile latency, ranging approximately between 3.3s and 3.5s. 
Also, enrichment with an external Redis cache performed comparably bad with a volatile latency and, oftentimes, being slower than asynchronous enrichment without cached database entries. 
The reason for the highly fluctuating latency may be the overhead incurred by the additional asynchronous operator and associated network connections, along with the implementation of the asynchronous Redis client.
Although latency is poor compared to the other enrichment methods in this evaluation, an external cache has the advantage over local caches of not being affected when \textit{TaskManagers} fail and does not need to be refilled. 
In order to show how a streaming job with a local cache behaves in the case of a failure of all \textit{TaskManagers}, we conducted an experiment in which all \textit{TaskManager} processes were deleted every few minutes. 
We chose the streaming job with the enrichment method with local cache and preceding custom partitioner for the experiment. 
~\autoref{fig:cache_partition_failures_latency} shows how the latency behaves in the case of a failure of all \textit{TaskManagers} at once. 
It can be seen that in the beginning, after restarting the streaming job at minutes 4 and 10, the volatility is comparatively high and then decreases after a short time and the latency becomes constant again. 
This is due to the fact that the cache must first be filled again by database accesses. 
~\autoref{fig:cache_partition_failures_cachehit} shows the corresponding cache hit rates. 
It can be seen that after all \textit{TaskManagers} fail, the cache hit is back at 100\% after a short time. 
This is because a single local cache contains only 3.000 entries and with a throughput of 4.000 events/s it is filled again very timely.
Note that our chosen latency measurement does not account for waiting time of events at the streaming platform, which is why we do not see a spike of latency in~\autoref{fig:cache_partition_failures_latency} after complete failure even though real consumer lag is experienced.

\subsubsection{State}

As an alternative to enriching events using a cache, we also evaluated data stream enrichment using Flink’s HashMapStateBackend.
Latency is measured over a period of 70 minutes, with a fixed throughput of 4.000 events/s.
We evaluated the enrichment method using two different amounts of historical transaction events, 2,000 and 200,000, since the amount of data managed within the Flink cluster has an impact on latency.
This means that depending on the execution, the available amount of enrichment data is written to Kafka and is then read by Flink as another data stream to enrich the current events.
The results, illustrated in~\autoref{fig:stream_enrichment}, showed that as the amount of enrichment data increased, the latency also increased. 
However, both methods of enrichment using the state backend had lower latency and minor fluctuations compared to enrichment using asynchronous database queries. 
Increasing the amount of data could potentially result in increased latency, and a geographically distributed Flink cluster could further increase network overhead and latency.

\subsection{Log Analytics Use Case}

\begin{figure}
 \begin{minipage}{\linewidth}
    \centering
    \includegraphics[width=\columnwidth]{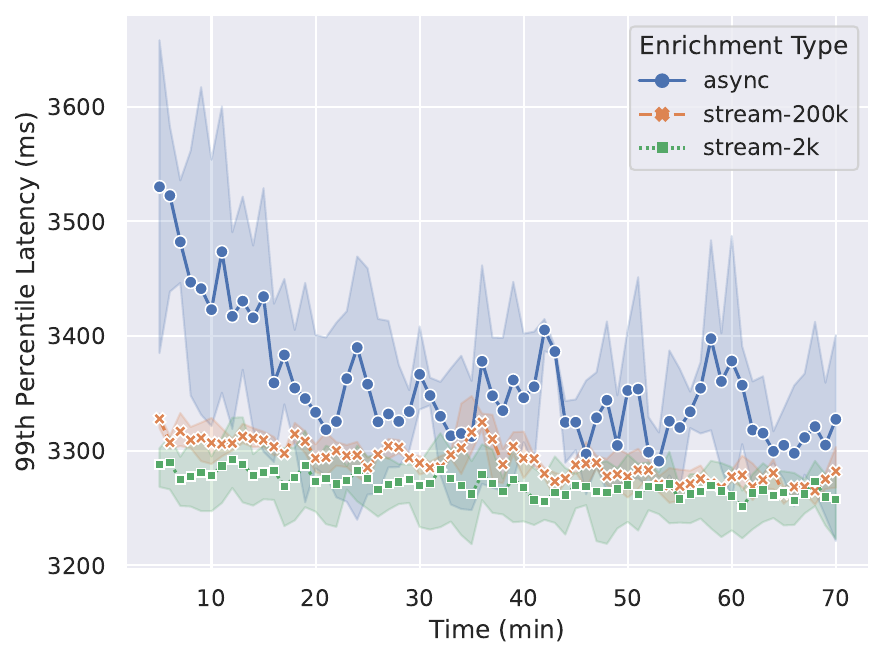}
    \caption{Comparing the latency of the state enrichment method with two different data amounts against enrichment by asynchronous database queries.}
    \label{fig:stream_enrichment}
 \end{minipage}
 
 \vspace{0.25cm}
 \begin{minipage}{\linewidth}
    \centering
    \includegraphics[width=\columnwidth]{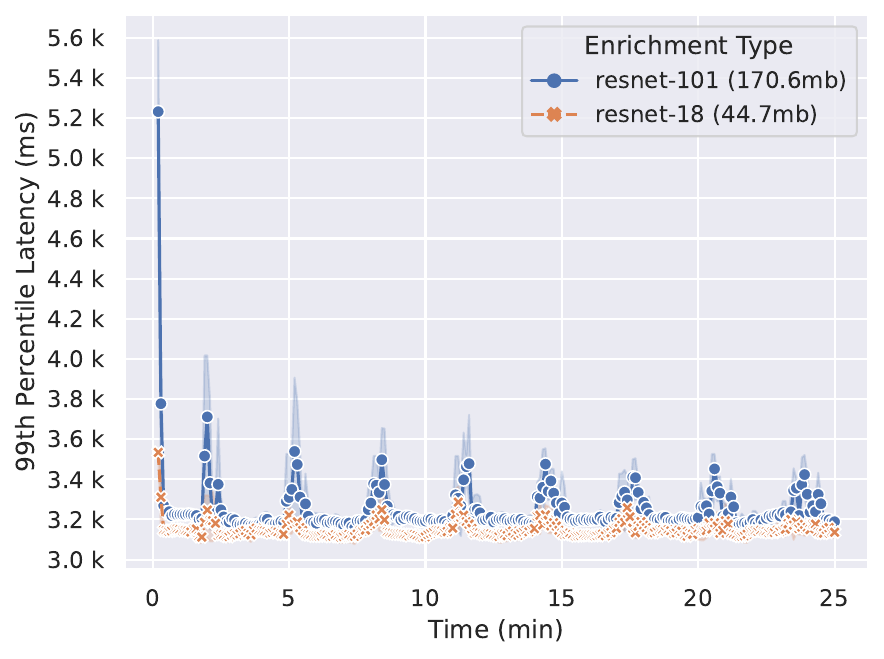}
    \caption{Latencies of the embedded executions of the two ResNet models. Every 4 minutes, the local cache containing the ONNX sessions was cleared to test the same worst-case scenario for each execution. The latency spikes in these periods can be attributed to the creation of the ONNX sessions.}
    \label{fig:model_time_latency}
 \end{minipage}
\end{figure}

For the log analytics use case, we evaluated the performance of embedded ML models in Flink with a focus on the number and size of models. 
The evaluation used pre-trained ResNet models, trained on the ImageNet dataset and provided by the ONNX community. 
ResNet models from the image classification domain were selected since different models are provided in terms of memory capacity. The context of the models does not fit into the area of log analytics, but the goal of this evaluation is to achieve a realistic performance analysis of embedded models.
The selected models were ResNet-18 with 18 layers and 44.7 MB memory size and ResNet-101 with 101 layers and 170.6 MB memory size. 
These models receive mini-batches of 3-channel RGB images of the form (N x 3 x H x W), where N is the batch size and H and W must be at least 224.
During the evaluation, the same image with a batch size of 1 and 224 for H and W was always selected as input. 
The models were stored in a local cache in Flink, which was flushed every four minutes to control the number of cache misses and create the same conditions for all executions. 
The throughput chosen for the evaluation was 1000 events/s, and the streaming job was run for 25 minutes with only 12 different keys that were normally distributed. 
For each key, the same model was used, which meant that when all caches were filled, the same model was cached 12 times.

The evaluation revealed that the Java Virtual Machine (JVM) Garbage Collector (GC) was slow in freeing the memory occupied by the ONNX session, which has a similar memory size as the models. 
As a result, the memory resources of a \textit{TaskManager} were quickly used up if new ONNX sessions were added several times per second, while removing old ones from the cache. 
This led to the \textit{TaskManager} crashing, and the entire streaming job had to be restarted. 
It was assumed that the amount of memory of the models was larger than the memory capacity of the \textit{TaskManagers}. 
To prevent the GC from being overwhelmed, the caches were flushed every four minutes, and the number of cache misses was controlled. 
The latencies of the embedded executions of the models were affected by the loading of the models from Google Cloud Storage and the subsequent creation of the ONNX session. 
The latency was highest at the beginning of the execution due to these processes.
The evaluation showed that the ONNX session creation time for the ResNet-18 model was almost up to half a second, while for the ResNet-101 model, it could take over a second. 
The prediction duration for the ResNet-101 model was also more than twice as long as that of the ResNet-18 model. 
The maximum execution times of these two processes for both models are shown in Table \ref{table:model_time_details}. 
The latencies of the ResNet-101 model, as depicted in~\autoref{fig:model_time_latency}, were higher than those of the ResNet-18 model, particularly during the periods when the cache was cleared every four minutes, increasing up to 3.6 seconds during these periods after the latencies normalized in the first few minutes.
Generally, we conclude that while good latencies can be obtained using embedded ML models, established frameworks such as Flink are not yet optimized for such usage, as indicated by our previous findings.

\subsection{Discussion}

\begin{table}
\centering
\caption{Model Details for Log Analytics Use Case}
\begin{tabular}{ccccc} 
 \toprule
 Model & Size & \shortstack{GCS\\Fetch Time} & \shortstack{Session\\Creation Time} & \shortstack{Prediction\\Time} \\
 \midrule
 resnet-18 & 44.7Mb & 871ms & 461ms & 90.8ms \\
 resnet-101 & 170.6Mb & 1663ms & 1098ms & 216.75ms \\
 \bottomrule
\end{tabular}
\label{table:model_time_details}
\end{table}

We presented the evaluation results of different enrichment methods executed in a real data infrastructure in GCP using Flink as a representative DSP system. 
These enrichment methods were to varying extents tested with two representative use cases, where the fraud detection use case can be associated with our defined categories of \emph{simple} and \emph{complex queries}, and the log analytics use case sheds light on a special case from category \emph{finite data sources}.
The experiments conducted to assess synchronous and asynchronous Cassandra queries showed that streaming jobs with asynchronous queries have better performance and are more resource-efficient. 
The evaluation of caching enrichment methods revealed that, for the investigated workloads, using a local cache rather than an external cache leads to better performance, but places a greater load on the resources of the respective DSP system. 
It should be noted that outsourcing data to a cache is not always possible due to the complexity of the database queries or the data structure. 
The use of state backend for enrichment demonstrated reliable enrichment of stream events, but the performance decreases with an increasing amount of enrichment data. 
Furthermore, running embedded ML models in Flink is not suitable for performance-heavy workloads in a single task, and multiple memory-intensive workloads tend to consume too much of available resources, causing the JVM Garbage Collector to free memory slowly.
All in all, and regardless of the details of the implementations, the results presented allow an assessment of in which case and at which approximate throughput rates certain methods of data enrichment are appropriate and what advantages and disadvantages they bring.
\section{Related Work}
\label{sec:related_work}

This section discusses related works on unifying batch and stream jobs, execution of ML models in and out of DSP systems, joining of stream and disk-based data, and event enrichment in modern DSP systems with disk-based databases. 
Its purpose is to contextualize the contribution of this work in enriching events in a DSP system with data from disk-based databases or generated data from a model.
\\
\textbf{Unified Batch and Stream Applications.} 
Service decoupling and the complexity of modern end-to-end data pipelines lead to an increasing overhead that may negatively impact performance. 
Arcon~\cite{MeldrumSKC0H19} and Neptune~\cite{GarefalakisKP19} address the unification of stream and batch processing to increase performance by providing an optimized common intermediate representation and dynamically prioritizing latency-critical jobs in unified stream and batch applications, respectively. 
While modern DSP systems offer the unified execution of stream and batch jobs, they cannot keep up with the query capabilities and memory sizes of modern databases. 
Enriching events during execution in a DSP system with additional data from a database leads to the merging of fast data streams and slow disk-based databases. 
Frameworks such as Kafka could be used in combination with a DSP system as data storage, but this is only applicable for a small number of use cases, as the query capabilities in Kafka are fairly limited compared to traditional databases. 
Therefore, there are no specific approaches to unifying large-scale data storage and latency-critical streaming applications to achieve more effective resource utilization and improved latency.\\
\textbf{Model Performance Evaluation.} 
The first performance evaluation study of model-serving integration tools in stream processing frameworks has been conducted in~\cite{HorchidanKKC22} by assessing the internal and external execution of a model in DSP systems. 
The integration of ML models assumes that the DSP system requires multiple models, which may exceed its storage capacity.
Additionally, the study considers different model sizes and addresses associated memory concerns. 
The work demonstrates that there are benefits to using integrated execution over external execution DSP frameworks, and that certain model formats offer superior performance.\\
\textbf{Data Warehouse Source Updates.}
Earlier works addressing the combination of high-speed data streams and slow disk-based databases involve active data warehousing. 
In this scenario, a data stream refers to quickly incoming events of source updates, which the data warehouse must process in real-time. 
One of the initial solutions to this issue is the MESHJOIN~\cite{PolyzotisSVSF08} algorithm, which has several variations and extensions~\cite{NaeemDWA10,NaeemDW12}. 
MESHJOIN fuses a high-speed data stream with a disk-based relationship, under the constraint of limited memory, using a hash-join. 
The algorithm scans the entire disk-based relationship sequentially at high speed, and the incoming events from the stream are processed in windows and then combined with the entries from the relationship. 
This approach distributes the expenses of the input-output operations across windows of stream events.\\
\textbf{Disk-based Database Enrichment.}
In~\cite{DerakhshanSS13}, an operator for DSP systems is proposed that enriches incoming stream events using a cache with data from a relational and disk-based database in a single node.
Depending on whether an incoming event causes a cache hit or a cache miss, it is processed in a thread for the respective category. 
Events causing cache misses are combined into batches and then used to query the database. 
Meanwhile, events causing cache hits are processed in parallel. After processing, the two sets of events are merged back into their original order. 
The paper reports higher throughput with this approach compared to a record-at-a-time approach. 
The experiments were conducted using a simulated DSP system and a MySQL database.
In~\cite{JeonLK19}, the authors describe a join between stream and disk-based data using a micro-batch model built on Spark Streaming~\cite{ZahariaDLHSS13} and MongoDB. 
The approach considers distributed execution of operators and assumes that the external disk-based data volume is larger than the storage capacity of the DSP system. 
To minimize database access, a cache is implemented in Spark that stores database entries in their own RDDs. 
If data is unavailable in the cache, a query is generated for multiple cache miss keys to reduce the number of queries. 
The authors also implemented a load-balancing mechanism by dynamically adjusting cache sizes in the DSP system to regulate the database load. 
In~\cite{KimL20c}, the authors extended~\cite{JeonLK19} to support similarity joins.
\section{Conclusion}
\label{sec:conclusion}

This paper conducted an assessment of various data enrichment methods for DSP systems. 
To account for a broad range of practical scenarios, three categories of data enrichment use cases were identified, covering database queries of different complexity as well as the embedding of state, and corresponding enrichment methods were designed and executed in Apache Flink.
To realistically evaluate these methods, representative use cases were implemented and run in a data infrastructure in a public cloud.
Our results showcase the advantages and limitations of the investigated methods, underlining the necessity for both sophisticated caching strategies for conventional data as well as better integration of embedded ML models into existing or future DSP systems.

To build on the work presented in this paper, future research could focus on building a dynamic caching mechanism that adapts to the access speeds of sources and different workloads.
Another promising direction is to explore methods that are reflecting more specialized situations, such as using GPU acceleration or cloud-based methods like serverless.

\section*{Acknowledgments}
This work has been supported through grants by the Google Cloud Research Credits program (award GCP211695187).

\bibliographystyle{IEEEtran}
\bibliography{bib}

\end{document}